\documentclass[twocolumn,amsmath,amssymb,showpacs,pra]{revtex4}
\usepackage{graphicx}% Include figure files
\usepackage{dcolumn}% Align table columns on decimal point
\usepackage{bm}% bold math

\newcommand{\rme}{\mathrm{e}}

\begin{document}

\preprint{APS/123-QED}

\title{Interpreting a nested Mach-Zehnder interferometer with classical optics}% Force line breaks with \\

\author{Pablo L. Saldanha}\email{saldanha@fisica.ufmg.br}
 %\altaffiliation[Also at ]{Physics Department, XYZ University.}%Lines break automatically or can be forced with \\
\affiliation{Departamento de F\'isica, Universidade Federal de Minas Gerais, Caixa Postal 701, 30161-970, Belo Horizonte, MG, Brazil}

%\author{Charlie Author}
% \homepage{http://www.Second.institution.edu/~Charlie.Author}
%\affiliation{
%Second institution and/or address\\
%This line break forced% with \\
%}%

\date{\today}% It is always \today, today,
             %  but any date may be explicitly specified

\begin{abstract}
In an recent work with the title ``Asking Photons Where They Have Been'', Danan \textit{et al.} experimentally demonstrate an intriguing behavior of photons in an interferometer [Phys. Rev. Lett. \textbf{111}, 240402 (2013)]. In their words: ``The photons tell us that they have been in the parts of the interferometer through which they could not pass.'' They interpret the results using the two-state vector formalism of quantum theory and say that, although an explanation of the experimental results in terms of classical electromagnetic waves in the interferometer is possible (and they provide a partial description), it is not so intuitive. Here we present a more detailed classical description of their experimental results, showing that it is actually intuitive. The same description is valid for the quantum wave function of the photons propagating in the interferometer. In particular, we show that it is essential that the wave propagates through all parts of the interferometer to describe the experimental results. We hope that our work helps to give a deeper understanding of these interesting experimental results.
\end{abstract}

% PACS, the Physics and Astronomy
                             % Classification Scheme.
                             
                % 42.50.Ct 	Quantum description of interaction of light and matter; related experiments             
                % 42.50.Dv 	Quantum state engineering and measurements (see also 03.65.Ud Entanglement and quantum nonlocality, e.g., EPR paradox, Bells inequalities, GHZ states, etc.)
                % 42.50.Ar 	Photon statistics and coherence theory 
                % 42.50.-p 	Quantum optics (for lasers, see 42.55.-f and 42.60.-v; see also 42.65.-k Nonlinear optics; 03.65.-w Quantum mechanics)
                % 03.65.Ta 	Foundations of quantum mechanics; measurement theory (for optical tests of quantum theory, see 42.50.Xa)

%\keywords{Suggested keywords}%Use showkeys class option if keyword
      %42.50.Ar 	Photon statistics and coherence theory
      %03.65.Ud 	Entanglement and quantum nonlocality

                          %display desired
\maketitle

%$k_y/\sigma$ $\tilde{\Psi}_A$ $\tilde{\Psi}_B$

The wave-particle duality is one of the most intriguing features of quantum mechanics. Quantum entities may behave as particles, as waves or as a strange combination of these possibilities. The affirmation that a quantum entity, such as an electron or a photon, is either a particle or a wave will always imply in a contradiction with experiments. So we can say that these entities are not particles nor waves, but very strange ``things'' that we do not understand in an intuitive way. This duality is explicitly manifested, for instance, in delayed choice experiments \cite{weeler78,jacques07,jacques08} and in quantum erasers \cite{scully91,herzog95,durr98,walborn02}. Recently delayed choice experiments were performed with quantum beam splitters \cite{roy12,auccaise12,tang12,peruzzo12,kaiser12} following the proposal of Ref. \cite{ionicioiu11}, showing even more intriguing behaviors.

In an interesting recent work, Danan \textit{et al.} demonstrated another experiment in which the wave-particle duality plays an important role in the nonintuitive experimental results  \cite{danan13}. This experiment was inspired on recent discussions about the past of a quantum particle in an interferometer \cite{vaidman13}. In the experimental arrangement, there is an inner interferometer in one of the arms of a large interferometer \cite{danan13}. They demonstrated that even when the inner interferometer is adjusted to produce destructive interference to the direction of the output port of the large interferometer, a tilting of some mirrors in the inner interferometer affects the average detection position of photons at the exit of the large interferometer. It is in this sense that the authors say that  the photons ``have been in the parts of the interferometer through which they could not pass'', since the dependence of the average detection position of the photons on the tilting of the mirrors shows that the photons have been in that arm, while the destructive interference in the inner interferometer should imply that the detected photons could not have passed through that arm. They also showed that the tilting of some other mirrors in this same arm of the large interferometer does not affect the average detection position of the photons at the exit. Danan \textit{et al.} argue that this is a strange behavior, since the dependence on the tilting of some mirrors shows that the photons have been in that arm, while the independence on the tilting of the other mirrors would imply that they have not been in that arm. The authors provide an explanation using the two-state vector formalism of quantum theory \cite{aharonov64,aharonov90} and claim that this would be the most intuitive explanation of the phenomenon. They also discuss that an explanation in terms of classical electromagnetic waves is certainly possible, since the experiment was done with a strong laser beam, and provide a partial classical description of their results \cite{danan13}.

%In a very interesting recent work, Danan \textit{et al.} demonstrated another experiment in which the wave-particle duality plays an important role in the nonintuitive experimental results  \cite{danan13}. This experiment was inspired on recent discussions about the past of a quantum particle in an interferometer \cite{vaidman13}. They showed that the average detection position of photons at the exit of the interferometer depends on the tilting of some mirrors in one of the interferometer arms, but not on the tilting of other mirrors in the same arm. It is in this sense that the authors say that the photons ``have been in the parts of the interferometer through which they could not pass'', since the dependence on the tilting of some mirrors shows that the photons have been in that arm, while the independence on the tinting of the other mirrors would imply that they have not been in that arm. The authors provide an explanation using the two-state vector formalism of quantum theory \cite{aharonov64,aharonov90} and claim that this would be the most intuitive explanation of the phenomenon. They also discuss that an explanation in terms of classical electromagnetic waves is certainly possible, since the experiment was done with a strong laser beam, and provide a partial classical description of their results \cite{danan13}. 

Here we provide a more detailed classical description of the experiments of Ref. \cite{danan13}, showing that it is actually  intuitive. The same description is valid for the propagation of the wave function of the photons in the interferometer. In particular, we show that the wave (be it classical or quantum) must pass through both arms of the large interferometer to explain the experimental results, and that the fact that some mirrors affect the average photon detection position and some do not can be understood in terms of wave interference in a simple way. So we hope to give a contribution for a better understanding of the interesting results of Ref.  \cite{danan13} with this work. 

The experimental setup of Ref. \cite{danan13} is depicted in Fig. 1. A large interferometer, with entrance and exit beam splitters BS$_1$ and BS$_4$, has an inner (nested) interferometer with entrance and exit beam splitters BS$_2$ and BS$_3$ in one of its arms. The modulus of the reflection and transmission coefficients of BS$_1$ and BS$_4$ are $\sqrt{\frac{2}{3}}$ and $\sqrt{\frac{1}{3}}$ respectively, while the modulus of the reflection and transmission coefficients of BS$_2$ and BS$_3$ are $1/\sqrt{2}$. %For simplicity, we will include any phase of the beam splitters reflection coefficients in the path difference phases of the interferometers. 
The mirrors $A$, $B$, $C$, $E$ and $F$ vibrate around their horizontal axes, causing an oscillation of the vertical position of beam reaching the detector $D$. The amplitude of this oscillation is much smaller than the beam diameter, and each mirror vibrates at a different frequency. The detector $D$ is a quad-cell photodetector and the registered signal is proportional to the difference between the number of photons detected in the upper and lower cells. The power spectrum of the measured signal contains peaks at the vibration frequencies of the mirrors whose angular positions affect the beam position at $D$. Since the vibration frequencies are different, the influence of each mirror on the beam position at $D$ can be extracted from the data.

\begin{figure}\begin{center}
  % Requires \usepackage{graphicx}
  \includegraphics[width=8cm]{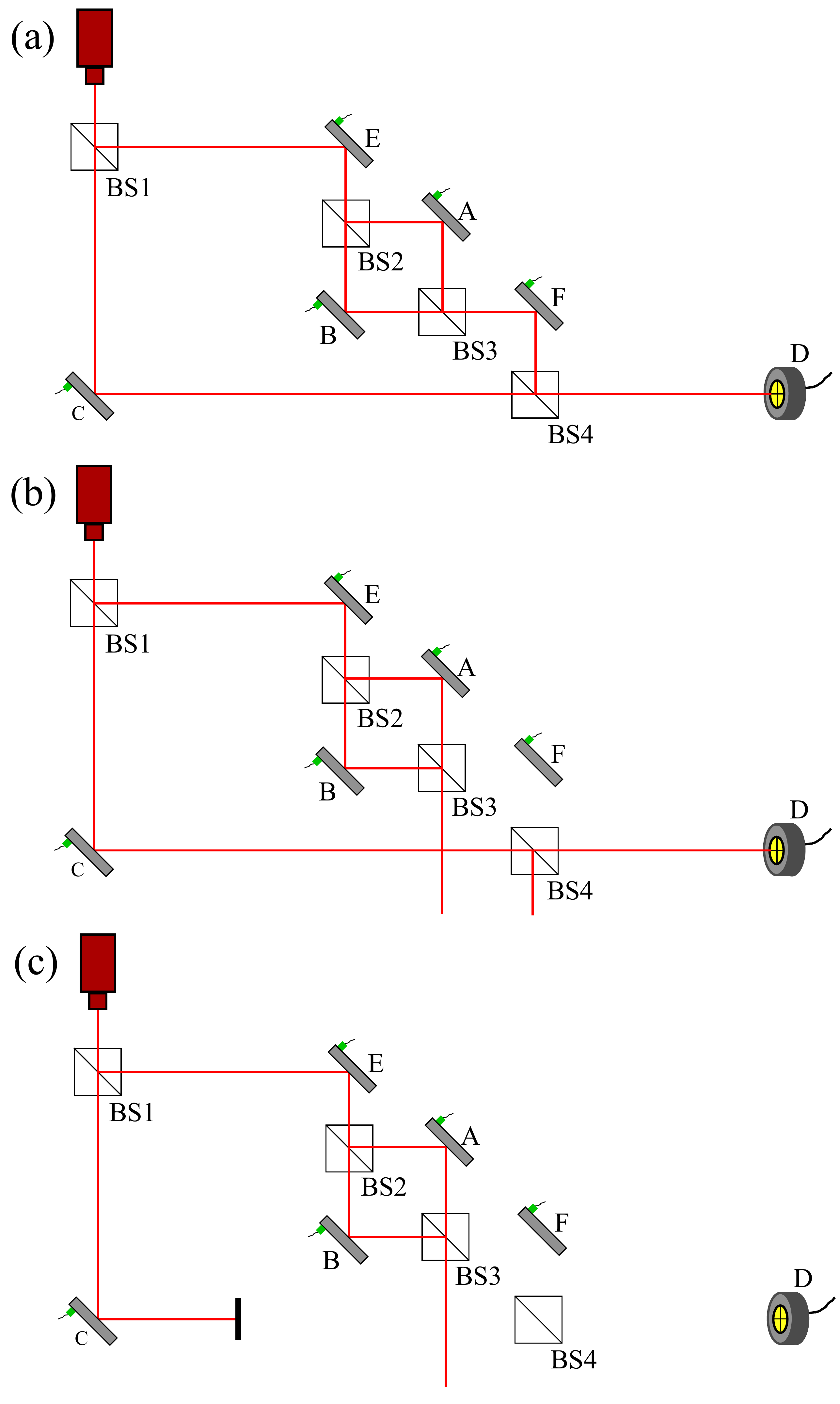}\\
  \caption{(Color online) Interferometer setup. A laser beam propagates through a complex interferometer and is detected by a quad-cell photodetector $D$ in one of its exits. The registered signal is proportional to the difference between the number of photons detected in the upper and lower cells. The large interferometer, with entrance and exit beam splitters BS$_1$ and BS$_4$, has an inner interferometer with entrance and exit beam splitters BS$_2$ and BS$_3$ in one of its arms. The mirrors $A$, $B$, $C$, $E$ and $F$ vibrate around their horizontal axes, causing an oscillation of the vertical position of beam reaching the detector. Each mirror has a different oscillation frequency and the power spectrum of the signal is obtained. (a) The inner interferometer is aligned to generate constructive interference to the direction of mirror $F$ and the large interferometer is aligned to generate constructive interference to the direction of the detector $D$. (b) The inner interferometer is aligned to generate destructive interference to the direction of mirror $F$. (c) The inner interferometer is aligned to generate destructive interference to the direction of mirror $F$ and the arm of mirror $C$ is blocked.}
\label{fig1}
 \end{center}\end{figure}

In the experimental situation of Fig. 1(a), the inner interferometer is aligned to generate constructive interference to the direction of mirror $F$ and the large interferometer is aligned to generate constructive interference to the direction of the detector $D$. The experimental results show that the angular position of all mirrors $A$, $B$, $C$, $E$ and $F$ affect the beam position at $D$ \cite{danan13}. The interesting and surprising result happens in  the situation of Fig. 1(b), where the inner interferometer is aligned to generate destructive interference to the direction of mirror $F$ and the large interferometer is aligned to generate constructive interference to the direction of the detector $D$. In this case the experimental results show that the angular positions of mirrors $A$, $B$ and $C$ affect the  beam position at $D$, but the angular positions of mirrors $E$ and $F$ do not \cite{danan13}. This result seems to imply that the photons interact with mirrors $A$ and $B$ without interacting with mirrors $E$ and $F$, what is, of course, impossible. In the experimental situation of Fig. 1(c), we have the same picture of Fig. 1(b) but the arm of mirror $C$ is blocked. In this case, the angular positions of none of the mirrors affect the  beam position at $D$ \cite{danan13}.

Let us start our classical description of the experiments with the situation of Fig. 1(b), which is the most interesting one. The photons come from a Gaussian laser beam, such that their amplitude can be written as  $\Psi(x,y,z,t)\propto\mathrm{e}^{-(x^2+y^2)/W^2}\cos(kz-\omega t)$ in a region close to its waist, where $\omega$ is the angular frequency of light and $k=\omega/c$ the modulus of the wavevector \cite{mandel}. This amplitude may refer to a component of the electric or magnetic field of the beam, or to a component of the real or imaginary part of the Bialynicki-Birula--Sipe wave function of the photons in the beam \cite{birula94,sipe95,saldanha11}. If the experiment is performed with electrons, neutrons or other massive particles, the amplitude would be of the real or imaginary part of the Schr\"odinger wave function of these particles. So our treatment is valid for a classical electromagnetic wave and for quantum particles.  Since the beam is in the paraxial regime, we can write its angular spectrum in terms of the components $k_x$ and $k_y$ of the wavevectors in the $xy$ plane as the Fourier transform of the amplitude \cite{mandel}, obtaining 
\begin{equation}\label{gauss}
	\tilde{\Psi}_T(k_x,k_y)=\tilde{\Psi}(k_x)\tilde{\Psi}(k_y),\;\;\mathrm{with}\;\;\tilde{\Psi}(k_i)=N\mathrm{e}^{-k_i^2/\sigma^2},
\end{equation}
with $\sigma=2/W$ and $N$ the normalization factor of the corresponding wave. On this way, the beam amplitude can be written as 
\begin{eqnarray}\nonumber
	\Psi(x,y,z,t)&\propto&\mathrm{Re}\Bigg[\int dk_x \int dk_y \tilde{\Psi}_T(k_x,k_y)\times\\
	&& \;\;\;\;\;\;\times\rme^{i(k_xx+ky_y+\sqrt{k^2-k_x^2-k_y^2}z-\omega t)}\Bigg],
\end{eqnarray}
in a decomposition in terms of the wavevectors, the paraxial regime meaning that $k_x\ll k$ and $k_y\ll k$ for all non-negligible values of $\tilde{\Psi}_T(k_x,k_y)$ \cite{mandel}. From now on we will consider only the $y$ dependence of the beam state, $\tilde{\Psi}(k_y)$, since the $x$ dependence does not change with the tilting of the mirrors.

Considering the $z$ direction as the propagation direction of the beam in each part of the interferometer for simplicity, if mirror $i$ is tilted, all wavevector components of the beam change with the reflection, such that the state of the beam after the reflection changes like $\tilde{\Psi}(k_y)\rightarrow\tilde{\Psi}_i=\tilde{\Psi}(k_y-\kappa_i)$, where $\tilde{\Psi}_i$ refers to the state of the beam just after mirror $i$ and $\kappa_i$ is proportional to the inclination amount. With this notation, the state of the beam just after mirror $F$ is
\begin{eqnarray}\label{psif}\nonumber
	\tilde{\Psi}_F=\frac{1}{\sqrt{6}}&&\big[ \tilde{\Psi}(k_y-\kappa_E-\kappa_A-\kappa_F) +\\
	                                      &&- \tilde{\Psi}(k_y-\kappa_E-\kappa_B-\kappa_F) \big].
\end{eqnarray}
In Fig. 2 we plot $\tilde{\Psi}_F$ for different values of $\kappa_A$, $\kappa_B$, $\kappa_E$ and $\kappa_F$. The vertical scale in the plots is arbitrary, but is the same in all plots. When $\kappa_A=\kappa_B$, as in Fig. 2(a), no light goes to mirror $F$. This is expected, since there is perfect destructive interference in the inner interferometer in this case. When $\kappa_A-\kappa_B>0$, as in Fig. 2(b), there are almost no wavevectors whose $y$ components are close to 0, since there is an almost perfect destructive interference. But for wavevectors whose $y$ components are close to $\pm\sigma$, we have some amplitude. This can be understood with the help of Fig. 3, that represents the beam states that come from mirrors $A$ ($\tilde{\Psi}_A$) and $B$ ($\tilde{\Psi}_B$) for exaggerated values of $\kappa_A$ (positive) and $\kappa_B$ (negative). When we compute $\tilde{\Psi}_F\propto\tilde{\Psi}_A-\tilde{\Psi}_B$ for the graphs of Fig. 3, it is clear that a curve like Fig. 2(b) is obtained.
%The reason is that there is small but non negligible difference between the amplitude of the beam that is reflected by mirror $A$ and the amplitude of the beam that is reflected from mirror $B$ for these components in this case, such that the interference is not perfectly destructive. This can be understood with the help of Fig. 3. If we subtract There is a phase difference of $\pi$ between the positive components of the wavevectors in the $y$ direction and the negative ones. 
When $\kappa_A-\kappa_B<0$, as in Fig. 2(c), there is an inversion of the positive and negative parts of the amplitude. The amplitude maximum is proportional to  $|\kappa_A-\kappa_B|$. The values of $\kappa_E$ and $\kappa_F$ almost do not change the amplitude of the wave that is reflected by mirror $F$, as can be seen comparing Figs. 2(b) and 2(d), that have the same values for $\kappa_A$ and $\kappa_B$ and different values for $\kappa_E$ and $\kappa_F$. There is no perceptible difference between Figs. 2(b) and 2(d). The reason is that the angular tilting of mirrors $E$ and $F$ does not change the interference behavior of the inner interferometer, they only displace the wavevectors distribution of the beam after mirror $F$. This displacement is very small, since in the plots of Fig. 2 and in the experiments of Ref. \cite{danan13} we have $\kappa_i$ at least 3 orders of magnitude smaller than $\sigma$, such that they do not significantly influence the state of the beam after mirror $F$.

\begin{figure}\begin{center}
  % Requires \usepackage{graphicx}
  \includegraphics[width=7cm]{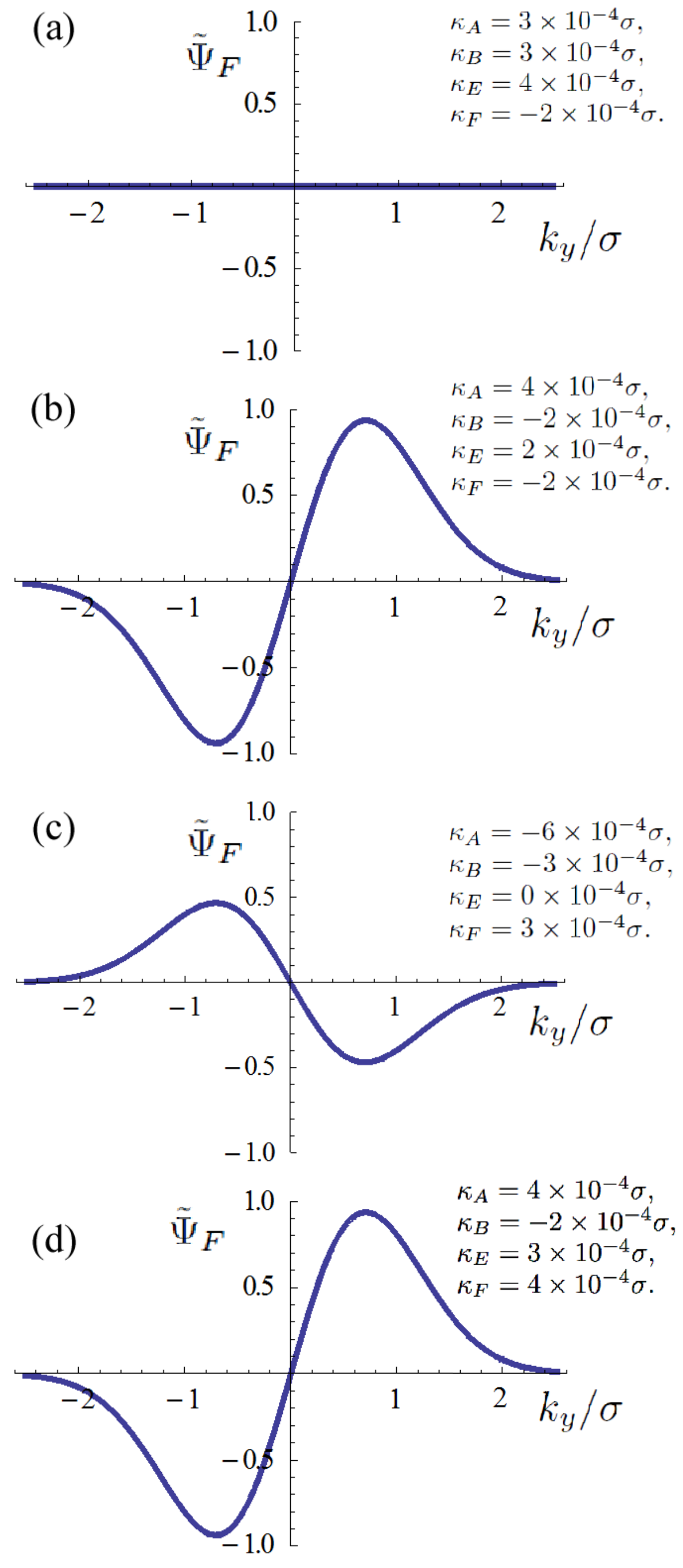}\\
  \caption{(Color online) Beam state after mirror $F$, given by $\tilde{\Psi}_F$ from Eq. (\ref{psif}), for different values of $\kappa_A$, $\kappa_B$, $\kappa_E$ and $\kappa_F$. 
  %(a) $\kappa_A=3\times10^{-4}\sigma$,  $\kappa_B=3\times10^{-4}\sigma$,  $\kappa_E=4\times10^{-4}\sigma$,  $\kappa_F=-2\times10^{-4}\sigma$. (b) $\kappa_A=4\times10^{-4}\sigma$,  $\kappa_B=-2\times10^{-4}\sigma$,  $\kappa_E=2\times10^{-4}\sigma$,  $\kappa_F=-2\times10^{-4}\sigma$. (c) $\kappa_A=-6\times10^{-4}\sigma$,  $\kappa_B=-3\times10^{-4}\sigma$,  $\kappa_E=0\times10^{-4}\sigma$,  $\kappa_F=3\times10^{-4}\sigma$. (d) $\kappa_A=4\times10^{-4}\sigma$,  $\kappa_B=-2\times10^{-4}\sigma$,  $\kappa_E=3\times10^{-4}\sigma$,  $\kappa_F=4\times10^{-4}\sigma$. 
  The vertical scale in the plots is arbitrary, but is the same in all plots. }
\label{fig2}
 \end{center}\end{figure}

%(a) 

%$\kappa_A=3\times10^{-4}\sigma$,  

%$\kappa_B=3\times10^{-4}\sigma$,  

%$\kappa_E=4\times10^{-4}\sigma$,  

%$\kappa_F=-2\times10^{-4}\sigma$. 

%(b) 

%$\kappa_A=4\times10^{-4}\sigma$,  

%$\kappa_B=-2\times10^{-4}\sigma$,  

%$\kappa_E=2\times10^{-4}\sigma$,  

%$\kappa_F=-2\times10^{-4}\sigma$. 

%(c) 

%$\kappa_A=-6\times10^{-4}\sigma$,  

%$\kappa_B=-3\times10^{-4}\sigma$,  

%$\kappa_E=0\times10^{-4}\sigma$,  

%$\kappa_F=3\times10^{-4}\sigma$. 

%(d)

%$\kappa_A=4\times10^{-4}\sigma$,  

%$\kappa_B=-2\times10^{-4}\sigma$,  

%$\kappa_E=3\times10^{-4}\sigma$,  

%$\kappa_F=4\times10^{-4}\sigma$.

\begin{figure}\begin{center}
  % Requires \usepackage{graphicx}
  \includegraphics[width=6.5cm]{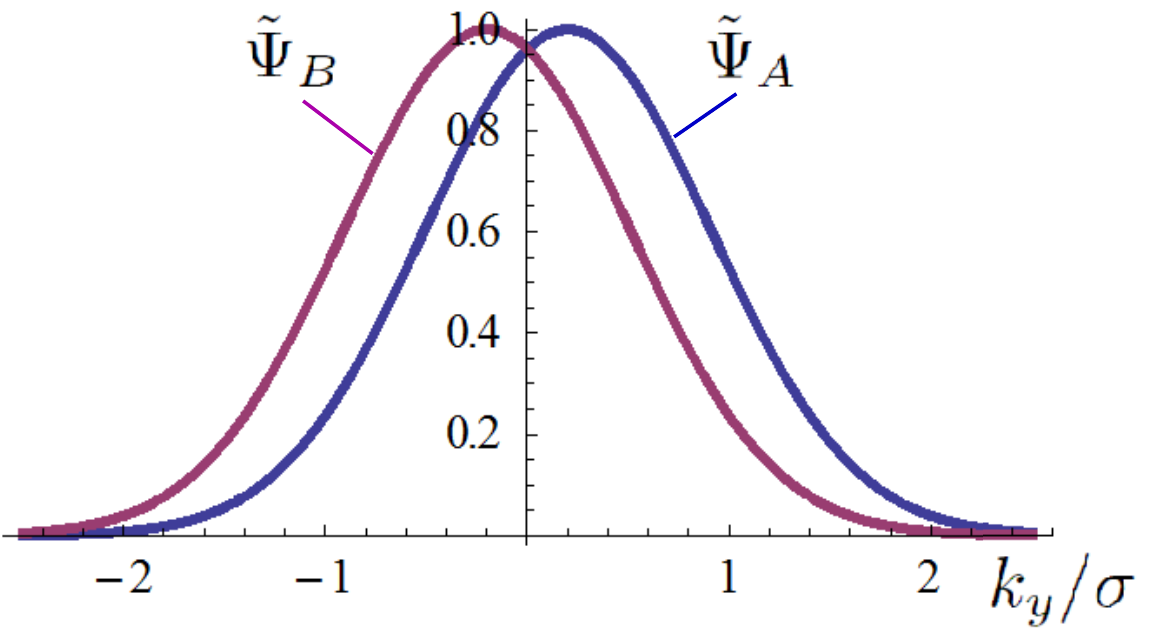}\\
  \caption{(Color online) Two Gaussian functions $\tilde{\Psi}_A$ and $\tilde{\Psi}_B$ with the same amplitude and width, but different centers.}
\label{fig3}
 \end{center}\end{figure}

The beam that goes in the direction of the detector $D$ is a superposition of the beam that comes from mirror $C$, with state $\tilde{\Psi}_C= \tilde{\Psi}(k_y-\kappa_C)/{\sqrt{3}}$, with the beam that comes from mirror $F$, with state given by $\tilde{\Psi}_F$ from Eq. (\ref{psif}). $\tilde{\Psi}_F$ has a much smaller amplitude than $\tilde{\Psi}_C$ due to the destructive interference in the inner interferometer. We can see that, when $\kappa_A>\kappa_B$, as in Figs. 2(b) and 2(d), the superposition of $\tilde{\Psi}_C$ with $\tilde{\Psi}_F$ will increase the components of $\tilde{\Psi}_C$ with positive $k_y$ and decrease the components with negative $k_y$. This results in a up displacement  of the resultant beam in the far field. When $\kappa_A<\kappa_B$, as in Fig. 2(c), the opposite happens and we have a down displacement of the resultant beam in the far field. And when $\kappa_A=\kappa_B$, as in Fig. 2(a), there is no displacement of the beam. Since $\kappa_E$ and $\kappa_F$ do not appreciably change the state $\tilde{\Psi}_F$ from Eq. (\ref{psif}), they do not affect the resultant beam displacement in the far field.

The amplitude of the resultant beam in the position of the detector $D$, considered to be in the far field, is proportional to the Fourier transform of the field amplitude in the interferometer, thus being proportional to the angular spectrum of the field at the exit of the interferometer \cite{mandel}. So it is proportional to $\tilde{\Psi}_D(k_y)=1/\sqrt{3}\,\tilde{\Psi}_C(k_y)+\sqrt{2/3}\,\tilde{\Psi}_F(k_y)$ with the substitution $k_y\rightarrow\alpha y$, $\alpha$ being a constant. The detector signal $\mathcal{S}$, being proportional to the difference between the light intensity in the upper and lower cells, is then proportional to the integral of the modulus squared of the angular spectrum of the resultant beam at the exit of the large interferometer for $k_y>0$ minus the same integral for $k_y<0$. So we obtain
\begin{equation}\label{signal}
	\mathcal{S}\propto\int_0^\infty |\tilde{\Psi}_D(k_y)|^2dk_y-	\int_{-\infty}^0 |\tilde{\Psi}_D(k_y)|^2dk_y
\end{equation}
with
\begin{eqnarray}\label{PsiD}
	&&\tilde{\Psi}_D(k_y)=  \frac{1}{3}\big[\tilde{\Psi}(k_y-\kappa_C) +\\\nonumber
	&&+ \tilde{\Psi}(k_y-\kappa_E-\kappa_A-\kappa_F)- \tilde{\Psi}(k_y-\kappa_E-\kappa_B-\kappa_F)\big].  
\end{eqnarray}
Since we are considering the limit $\kappa_i\ll\sigma$, we can write
\begin{equation}
	\tilde{\Psi}(k_y-\kappa_i)\approx \tilde{\Psi}(k_y)-\kappa_i\frac{\partial \tilde{\Psi}(k)}{\partial k}\Big|_{k=k_y}.
\end{equation}
Substituting this approximation for the terms on the right side of Eq. (\ref{PsiD}), we obtain
\begin{eqnarray}\label{PsiD2}\nonumber
	\tilde{\Psi}_D(k_y) &\approx&  \frac{1}{3}\Bigg[\tilde{\Psi}(k_y) - (\kappa_C+\kappa_A-\kappa_B) \frac{\partial \tilde{\Psi}(k)}{\partial k}\Big|_{k=k_y}\Bigg]\\
	                     &\approx& \frac{1}{3} \tilde{\Psi}(k_y-(\kappa_C+\kappa_A-\kappa_B)).
\end{eqnarray}
According to Eq. (\ref{gauss}), the function  $\tilde{\Psi}_D(k_y)$ from the above equation is symmetric around the maximum at $k_y=\kappa_C+\kappa_A-\kappa_B$. So, considering initially that we have $\kappa_C+\kappa_A-\kappa_B>0$, the integral of the first term on the right side of Eq. (\ref{signal}) from $2(\kappa_C+\kappa_A-\kappa_B)$ to $\infty$ cancels the second integral in this equation. In this case we have 
\begin{equation}\label{signal2}
	\mathcal{S}\propto\int_0^{2(\kappa_C+\kappa_A-\kappa_B)} |\tilde{\Psi}_D(k_y)|^2dk_y.
\end{equation}
Since the function $\tilde{\Psi}_D(k_y)$ is approximately constant on the above integral, we have
\begin{equation}
	\mathcal{S}\propto\kappa_C+\kappa_A-\kappa_B.
\end{equation}
It is easy to see that the above equation is also valid for $\kappa_C+\kappa_A-\kappa_B<0$. So when the inner interferometer causes a destructive interference for the light propagation in the direction of mirror $F$, as in Fig. 1(b), the vibrations of the mirrors $A$, $B$ and $C$ make the signal $\mathcal{S}$ to oscillate, while the vibrations of mirrors $E$ and $F$ do not. For this reason, if all mirrors vibrate at different frequencies, the power spectrum of the signal will present peaks at the vibration frequencies of mirrors $A$, $B$ and $C$, but not at the vibration frequencies of mirrors $E$ and $F$. This is shown in the experiments of Ref. \cite{danan13}.

Considering now the situation depicted in Fig. 1(a), where there is constructive interference in the inner interferometer for the photons to go out in the direction of mirror $F$, the calculations are basically the same. But now the angular spectrum of the resultant beam at the exit of the interferometer can be written as 
\begin{eqnarray}
	&&\tilde{\Psi}_D'(k_y)=  \frac{1}{3}\big[  \tilde{\Psi}(k_y-\kappa_C) +\\\nonumber
	&&+ \tilde{\Psi}(k_y-\kappa_E-\kappa_A-\kappa_F)+ \tilde{\Psi}(k_y-\kappa_E-\kappa_B-\kappa_F)\big].  
\end{eqnarray}
Substituting in Eq. (\ref{signal}) in the limit $\kappa_i\ll\sigma$ and following the same steps as before in the calculation, we obtain
\begin{equation}
	\mathcal{S}'\propto\kappa_C+\kappa_A+\kappa_B+2\kappa_E+2\kappa_F.
\end{equation}
Now the vibrations of all mirrors make the signal  $\mathcal{S}$ oscillate, such that the vibration frequencies of all mirrors should be in the power spectrum. Since the tilting  of mirrors $E$ and $F$ causes twice the beam displacement than a tilting of the same amount of the other mirrors, the power spectrum at the vibration frequencies of mirrors $E$ and $F$ should have a peak four times the value of the peaks at the vibration frequencies of mirrors $A$, $B$ and $C$, since the power spectrum is proportional to the square of the oscillation amplitude at each frequency. This is shown in the experiments of Ref. \cite{danan13}.

%Considering the situation depicted in Fig. 1(c), all light that goes to the detector comes from mirror $F$. We can see that, according to the plots in Fig. 2, the light intensity (which is proportional to the square of the curves in Fig. 2) is zero close to the origin up to the first order in the $\kappa_i$ for all curves.  So the vibration of the mirrors should not affect the signal $\mathcal{S}$, since small shifts of the intensity curves obtained from Fig. 2 do not change the difference between the integral of the intensity for $\kappa_y>0$ and for $\kappa_y<0$, such that no peak should be observed in the power spectrum. This is also shown in the experiments of Ref. \cite{danan13}.

Considering the situation depicted in Fig. 1(c), all light that goes to the detector comes from mirror $F$. We can see that, according to the plots in Fig. 2, the light intensity will be always (almost) symmetric around the origin, such that the vibration of the mirrors should not affect the signal $\mathcal{S}$ and no peak should be observed in the power spectrum. This is also shown in the experiments of Ref. \cite{danan13}.

To summarize, we have presented a description of the experiments of Ref. \cite{danan13} using a classical description of light propagating in the interferometer. The same description is valid for the propagation of the wavefunction of quantum particles in the interferometer. This interpretation, that clearly shows that it is essential that the wave propagates through all parts of the interferometer to describe the experimental results, is complementary to the one using the two-state vector formalism of quantum theory \cite{aharonov64,aharonov90} presented in  Ref. \cite{danan13}. We hope that our results can give more physical insight to the behavior of photons (and electromagnetic waves) in this interesting kind of interferometers.

This work was supported by the Brazilian agencies CNPq and PRPq/UFMG.

%\bibliography{refer}% Produces the bibliography via BibTeX.

\begin{thebibliography}{33}

\bibitem{weeler78} J. A. Wheeler, in \textit{Mathematical Foundations of Quantum Mechanics}, edited by A. R. Marlow (Academic, New York, 1978).

\bibitem{jacques07} V. Jacques \textit{et al.}, Science \textbf{315}, 966 (2007).

\bibitem{jacques08} V. Jacques \textit{et al.}, Phys. Rev. Lett. \textbf{100}, 220402 (2008).

\bibitem{scully91} M. O. Scully, B. G. Englert, and H. Walther, Nature \textbf{351}, 111 (1991).

\bibitem{herzog95} T. J. Herzog, P. G. Kwiat, H. Weinfurter, and A. Zeilinger, Phys. Rev. Lett. \textbf{75}, 3034 (1995).

\bibitem{durr98} S. D\"urr, T. Nonn, and G. Rempe, Nature \textbf{395}, 33 (1998).

\bibitem{walborn02} S. P. Walborn, M. O. Terra Cunha, S. P\'adua, and C. H. Monken, Phys. Rev. A \textbf{65}, 033818 (2002).

\bibitem{roy12} S. S. Roy, A. Shukla, and T. S. Mahesh, Phys. Rev. A \textbf{85}, 022109 (2012).

\bibitem{auccaise12} R. Auccaise \textit{et al.}, Phys. Rev. A \textbf{85}, 032121 (2012). 

\bibitem{tang12} J.-S. Tang \textit{et al.}, Nat. Photonics \textbf{6}, 602 (2012).

\bibitem{peruzzo12} A. Peruzzo \textit{et al.}, Science \textbf{338}, 634 (2012).

\bibitem{kaiser12} F. Kaiser \textit{et al.}, Science \textbf{338}, 637 (2012).

\bibitem{ionicioiu11} R. Ionicioiu and D. R. Terno, Phys. Rev. Lett. \textbf{107}, 230406 (2011).

\bibitem{danan13} A. Danan, D. Farfurnik, S. Bar-Ad, and L. Vaidman, Phys. Rev. Lett. \textbf{111}, 240402 (2013).

\bibitem{vaidman13} L. Vaidman, Phys. Rev. A \textbf{87}, 052104 (2013); Z. H. Li, M. Al-Amri, and M. S. Zubairy, Phys. Rev. A \textbf{88}, 046102 (2013); L. Vaidman, Phys. Rev. A \textbf{88}, 046103 (2013). 

\bibitem{aharonov64} Y. Aharonov, P.G. Bergmann, and J. L. Lebowitz, Phys. Rev. \textbf{134}, B1410 (1964).
 
\bibitem{aharonov90} Y. Aharonov and L. Vaidman, Phys. Rev. A \textbf{41}, 11 (1990). 

\bibitem{birula94} I. Bialynicki-Birula. Acta Phys. Pol. A \textbf{86}, 97 (1994). 

\bibitem{sipe95} J. E. Sipe, Phys. Rev. A \textbf{52}, 1875 (1995). 

\bibitem{saldanha11} P. L. Saldanha and C. H. Monken, New J. Phys. \textbf{13}, 073015 (2011).

\bibitem{mandel} L. Mandel and E. Wolf, \textit{Optical Coherence and Quantum Optics} (Cambridge University Press, New York, 1995).



\end{thebibliography}

\end{document}